\documentclass[amssymb,prl,twocolumn,floats]{revtex4}

\usepackage{bm}
\usepackage{graphicx}

\begin{document}
\preprint{August 20, 2007}

\title{Optical manipulation of a single Mn spin in a CdTe-based quantum dot}

\author{M. Goryca}
\email[]{mgoryca@fuw.edu.pl}
\author{T. Kazimierczuk}
\author{M. Nawrocki}
\author{A. Golnik }
\author{J.A. Gaj}
\author{P. Kossacki}

\affiliation{%
Institute of Experimental Physics, University of Warsaw,\\
Ho\.za 69, 00-681 Warsaw, Poland }%

\author{P. Wojnar}
\author{G. Karczewski}

\affiliation{%
Institute of Physics, Polish Academy of Sciences,\\
Al. Lotnik\'ow 32/64, 02-688 Warsaw, Poland}%

\date{\today}

\begin{abstract}

A system of two coupled CdTe quantum dots, one of them containing a single Mn ion, was studied in continuous wave and modulated photoluminescence, photoluminescence excitation, and photon correlation experiments. Optical writing of information in the spin state of the Mn ion has been demonstrated, using orientation of the Mn spin by spin-polarized carriers transferred from the neighbor quantum dot. Mn spin orientation time values from 20 ns to 100 ns were measured, depending on the excitation power. Storage time of the information in the Mn spin was found to be enhanced by application of a static magnetic field of 1 T, reaching hundreds of microseconds in the dark. Simple rate equation models were found to describe correctly static and dynamical properties of the system.

\end{abstract}

\pacs{71.35.Pq; 71.70.Gm; 75.50.Dd; 78.55.Et; 78.67.De; 85.75.-d}

\draft

\keywords{}




\maketitle


One of important research directions that may influence the future of information processing, especially of spintronics \cite{zutic04}, is focused on physical phenomena occurring in nanoscale-size quantum objects. One of such objects, close to the ultimate limit of information storage miniaturization, is a single Mn atom in a semiconductor quantum dot (QD) \cite{Bhattacharjee03,FernandezPRB}. After intensive studies of semimagnetic QD containing many magnetic ions \cite{Dorozhkin03,Mackowski05,Bacher05,Zaitsev06}, single Mn atoms in CdTe \cite{besombesPRL}  and InAs \cite{kudelskiPRL}  QDs have been observed in photoluminescence (PL) experiments. Many experiments supplied substantial knowledge on physical properties of single Mn atoms, especially in CdTe QDs. In particular, they revealed a strong influence of the position of the Mn atom in the QD, reflecting the symmetry of the system \cite{legerPRL05}. They demonstrated an efficient optical read-out of the Mn spin state \cite{besombesPRL}. Furthermore, the dynamics of this state has been studied in photon correlation experiments \cite{besombesPRB08}, revealing an important influence of photo-created carriers on Mn spin relaxation. The writing and storing of the information in the Mn spin state has received less attention so far. These issues represent the focus of the present work.

In particular, we demonstrate optical writing of information in the spin state of a single Mn ion and we test the stability of this state in the time range up to 0.2~ms.

\begin{figure}[t]
\begin{center}
\includegraphics[width=85mm]{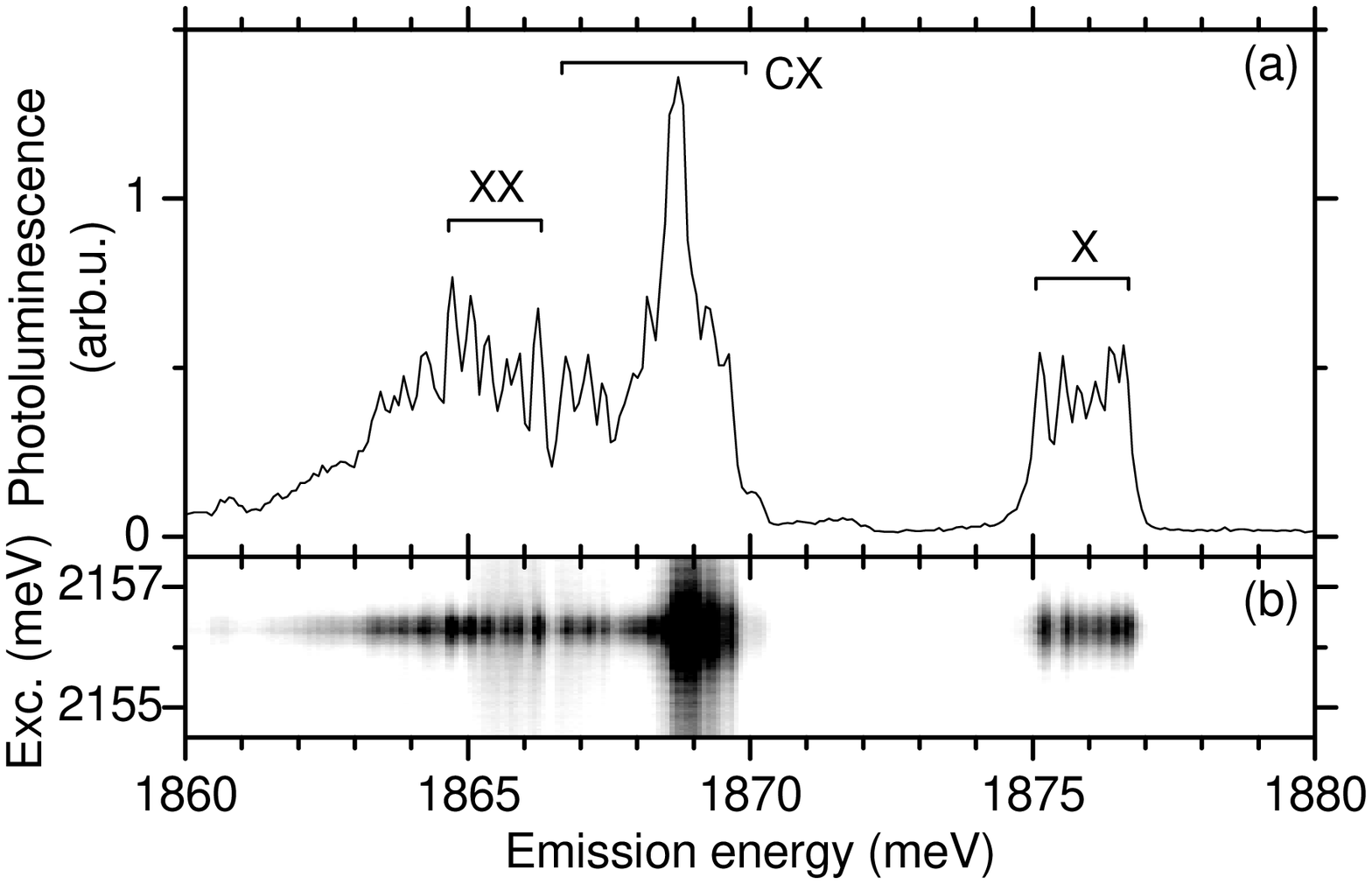}
\end{center}
\caption[]{ (a) Photoluminescence spectrum of a single QD with a single Mn ion, excited by linear polarization with indicated features related to neutral exciton (X), biexciton (XX), and charged excitons (CX); (b) Photoluminescence excitation map (density plot of PL intensity versus excitation/emission photon energy) close to the resonance.}
\label{spectra}
\end{figure}

CdTe QDs containing single Mn ions were grown by molecular beam epitaxy. A single layer of self-assembled QDs was deposited in a ZnTe matrix. Manganese was added by briefly opening the Mn effusion cell during the growth of the CdTe layer \cite{wojnar2007}. The opening time and the Mn flux were adjusted to achieve a large probability of growth of QDs with a single Mn ion in each dot. The selection of single QDs was done by spatial limitation of PL excitation and detection to an area smaller than 0.5 micrometer in diameter, with microscope objective immersed in pumped liquid helium. Continuous wave excitation was used either above the ZnTe barrier gap (at 457~nm) or by a tunable dye laser in the range 570 - 600~nm. Well separated photoluminescence lines from individual QDs were observed in the low energy part of the PL spectrum. We were able to select numerous lines showing a PL pattern characteristic for the presence of a single Mn ion in a dot \cite{FernandezPRB,besombesPRL}. An example spectrum is presented on Fig.~\ref{spectra}a. It contains six lines related to neutral exciton (X), equally spaced over a range of 1.5~meV. This splitting in six components is due to exciton-Mn exchange interaction. At energy about 11~meV below, biexciton (X) PL from the same QD is also split in 6 lines with the same splitting value. As a test of the identification of these two transitions, photon correlation measurements were performed in a Hanbury-Brown and Twiss setup \cite{HBT} with two monochromators equipped with avalanche photodiode detectors. The obtained biexciton-exciton cross-correlation histograms (not shown) exhibit a bunching peak related to a radiative cascade, confirming the identification of both transitions and their common origin (the same QD). Between the two features there are series of lines related to charged excitons. We performed systematic measurements of PL spectra for various values of excitation power and photon energy. The results were coherent with the behavior typical for similar QDs \cite{Suffczynski06}. Among the identified QDs with single Mn ions we selected those exhibiting a sharp resonance in photoluminescence excitation (PLE) spectra, as shown in Fig.~\ref{spectra}b.

Similar resonances were found previously in CdTe/ZnTe QDs without Manganese and interpreted as related to transfer of excitons between a smaller and a larger QD \cite{kazimierczuk2008}. Following arguments were used in support of the inter-dot energy  transfer: (i) resonances are very sharp; (ii) they appear at energy much larger than energy of observed emission (more than 0.2~eV), with no correlation between the two energies; (iii) all the observed charge states of the emitting QD share the same resonance energy; (iv) optical in-plane anisotropy (characteristic for the neutral exciton) is different and uncorrelated in absorption and in emission. We checked that all the above features occur in our case, except of anisotropy properties, since no fine structure splitting was detected. Additionally, we found that while the PL exciton lines exhibit characteristic sixfold splitting due to the interaction with single Mn ions, the absorption line is not split by this interaction (see Fig. 1b). Basing on these findings we conclude that the resonance is related to absorption in a QD with no Manganese. The photocreated exciton is then transferred to a larger dot containing a single Mn ion. 

\begin{figure}[t]
\begin{center}
\includegraphics[width=85mm]{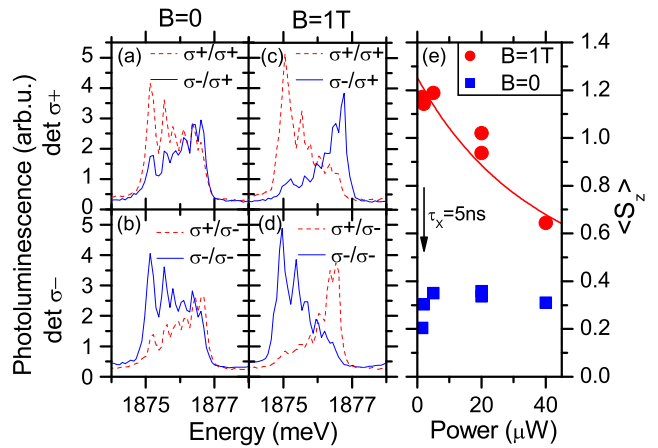}
\end{center}
\caption[]{ (a-d) Photoluminescence spectra excited at the resonance, using indicated excitation/detection polarizations. Measurements taken at T~=~1.5~K, and B~=~0~T (a-b) or B~=~1~T (c-d). (e) The mean spin of the Mn ion as a function of excitation power for B~=~0 and B~=~1~T compared with model described in text (solid line). The horizontal arrow indicates the power at which averaged time delay between capture of two excitons ($\tau_{X}$) is equal to 5~ns.}
\label{optical}
\end{figure}

Similarly as in Ref. \onlinecite{kazimierczuk2008}, the transferred exciton conserves its spin. The efficiency of the polarization transfer depends on details of the parameters of coupled QDs (such as in-plane anisotropy and excitation transfer rate). In our case the efficiency is almost independent of magnetic field, and decreases for higher excitation power due to the increased probability of the biexciton formation. The presence of biexcitons leads to a suppression of  the spin polarization transfer, since the biexciton singlet ground state cannot carry any spin memory. 

The spin transfer is clearly visible on Fig.~\ref{optical}a-d. The exciton PL spectra measured in the same circular polarizations of excitation and detection (co-polarized) have higher integrated intensity than those measured in different polarizations (cross-polarized). The lower intensity equals about 65~\% of the higher one for excitation power of about 6~$\mu$W. 

The exciton polarization transfer is used as a tool for optical writing of information in the Mn spin. Carriers created by defined circular polarization of light act on the Mn ion via exchange interaction and orient its spin. The Mn spin orientation appears in Fig.~\ref{optical}a-d as a nonuniform distribution of intensities between the 6 exciton lines. As the Mn spin polarization results from interaction with spin-polarized carriers, the spin polarizations of both species are related to each other. This relation will be discussed later. As a quantitative measure of the spin orientation we use the mean spin value of the Mn ion determined from the "center of mass" position of the PL sextet. An example variation of this value versus light power is shown in Fig.~\ref{optical}e (B~=~0 data). It shows a maximum at about 0.1~mW/$\mu$$m^2$ and decreases for higher and lower power. The decrease at the high power corresponds to the increased probability of the biexciton formation and - as a consequence - a decrease of the efficiency of the Mn spin alignment. The decrease at low excitation power is due to the competition between the Mn spin relaxation, and the spin orientation by the photocreated excitons. The relaxation becomes dominant when the average time delay between capture of two excitons is longer than the relaxation time. This condition provides an estimate of the relaxation time, being in our case of the order of a few nanoseconds. This value compares reasonably with the relaxation time determined for very diluted (Cd,Mn)Te quantum wells in absence of magnetic field \cite{Mateusz2008}. The fast relaxation was caused mainly by hyperfine interaction with nuclear spin.

As shown in Ref. \onlinecite{Mateusz2008}, the fast relaxation may be blocked by applying a small magnetic field, which suppresses mixing of Mn spin eigenstates by the hyperfine interaction. Therefore in order to define pure spin states of the Mn ion, a magnetic field of about 1~T was applied. The field was weak enough to assure negligible orientation of the Mn spin by its thermal distribution. This was checked when QD was excited by linearly polarized light. The mean spin was then equal about 0.2. In contrast, under circularly polarized excitation the mean spin was up to 1.2. The variation of the mean Mn spin versus excitation power at a fixed magnetic field and polarization of the light is shown in Fig.~\ref{optical}e. Similarly to the zero-field case, it shows a decrease at high power, resulting from an increase of the probability of biexciton formation. However, no decrease is observed at low power, down to its lowest values. This might be seen as an argument for a long spin relaxation time. 

The low excitation power limit in magnetic field corresponds to a particularly simple situation, when the only mechanism influencing the Mn spin state is the interaction with photocreated carriers present in the emitting dot. A simple optical orientation model can be tested in this situation. It is based on the assumption that any Mn state, characterized by its quantum number $m$, can be transferred to a state with $m$ different by 1 with a fixed probability $p_0$ during the presence of a suitably polarized exciton in the QD. Within the assumptions of the model, the steady-state distribution of Mn states depends only on the relative numbers of spin-up and spin-down excitons arriving in the emitting dot, and not on absolute exciton transfer rate or $p_0$ value. To test the applicability of the model, we sum the up- and down Mn spin transfer probabilities weighted with the distribution of Mn states read from the sextuplet spectrum of Fig.~\ref{optical}c-d and relative frequencies of arriving spin-up and spin-down excitons in the dot, known from the previously determined exciton spin transfer efficiency. The up- and down transfer rates thus obtained are equal with good accuracy, as required in a steady state. In other words, the exciton polarization transfer efficiency is directly correlated with the Mn spin state. 

The model can be easily extended to take into account the influence of biexcitons, assuming that each of them contributes with equal probability $p_0$/2 to the Mn up- and down spin flips. To determine the biexciton/exciton ratio of arrival rates in the QD, we use a rate equation model of exciton dynamics introduced by Regelman et al. \cite{regelman01}. We select a basis containing up to three excitons in the dot. Using thus obtained biexciton/exciton arrival rate ratio, power dependence of Mn spin orientation was simulated, with exciton spin transfer efficiency as a free parameter. An excellent agreement was obtained (Fig. 2e), with spin-up to spin-down probability ratio of 0.62. This value agrees well with that determined experimentally from net polarization of the exciton line at low excitation power.

\begin{figure}[t]
\begin{center}
\includegraphics[width=85mm]{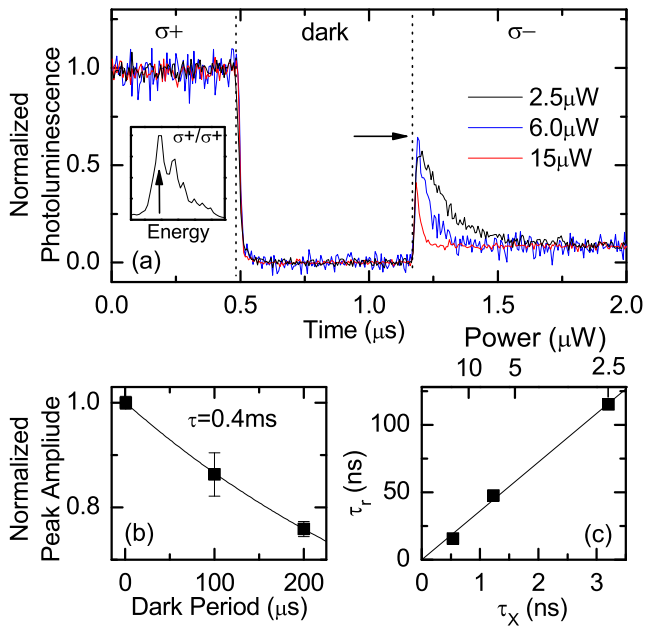}
\end{center}
\caption[]{ (a) Temporal profile of the PL intensity at one of the six components of the exciton spectrum (indicated in inset) during the excitation sequence described in text for B~=~1~T and indicated values of excitation power. The arrow indicates the ratio between PL intensity measured in co- and cross-polarization setup at excitation power equal to 6~$\mu$W. (b) PL intensity after turning on the excitation in $\sigma^{-}$ polarization vs. length of dark period. (c) The Mn reorientation time upon illumination in $\sigma^{-}$ polarization ($\tau_{r}$) vs. averaged time delay between capture of two excitons ($\tau_{X}$) and excitation power.}
\label{dynamics}
\end{figure}

The spin relaxation time was analyzed in more detail in an experiment, in which the excitation and its polarization were modulated. This was achieved by passing the laser beam through acousto-optic and electro-optic modulators. Using the first one we were able to switch the excitation on and off. The second modulator allowed us to change the circular polarization of the laser. The switching time of both modulations was 10~ns. The modulators were driven by a set of generators synchronized with a time-resolved photon counting system, which was used to record the temporal profiles of the PL signal. The excitation sequence was as follows: first the Mn spin state was set by $\sigma^{+}$ - polarized excitation. Then the light was switched off for a controlled delay and subsequently the light with opposite circular polarization was switched on to accomplish a read-out of the spin state. This pattern was repeated with a sufficiently low frequency to assure full setting of the the spin state each time the light was on. The Mn state was read by measuring PL intensity of a selected component of the exciton sextuplet in a given circular polarization. Its evolution reflected the orientation of the Mn spin combined with the exciton optical orientation. The measurements were repeated for both circular polarizations of detection. An example temporal profile is presented on Fig.~\ref{dynamics}a. It was obtained with the detection polarization $\sigma^{+}$ and detection energy set at the lowest component of the exciton sextuplet. 

Several features present in the profile may be used for determination of different properties of the spin orientation. First flat region of the profile is related to excitation with $\sigma^{+}$ polarization, the same as the detection polarization. When the excitation is switched off, the PL signal drops to zero and remains vanishing as long as the excitation is off. When subsequently a $\sigma^{-}$ polarized excitation is turned on, the PL intensity first rapidly rises up to an initial value determined by the Mn spin state and the efficiency of the exciton polarization transfer. If the Mn orientation is conserved, this value is given only by the exciton polarization transfer and should be equal to about 65\% of the previous value for excitation power equal to 6~$\mu$W (as marked in the Fig.~\ref{dynamics}a). Any loss of the Mn spin orientation would decrease this value. Fig.~\ref{dynamics}a shows clearly that the Mn spin state directly after the dark period was very close to that written by the light. The initial value of PL intensity is presented versus delay in Fig.~\ref{dynamics}b. Its decrease allows us to roughly estimate the Mn spin relaxation time in the dark to be about 0.4~ms  

After the initial rapid increase the PL intensity slowly decreases with a characteristic time of tens of nanoseconds. This decrease is related to reorientation of the Mn spin. The asymptotic intensity levels for excitation in the two polarizations are different because of different orientation of Mn spin and of the exciton polarization transfer. In the configuration selected for measurements presented in Fig. 3 both contributions combine with the same sign, producing an enhancement of PL intensity for excitation in $\sigma^{+}$, and its suppression for $\sigma^{-}$ excitation.   

The orientation of the Mn spin by circularly polarized light accelerates with increasing excitation power as shown in Fig.~\ref{dynamics}a. The characteristic time of this process is approximately inversely proportional to the excitation power. This property suggests that the simple model, introduced to describe the steady-state Mn spin orientation, can be used also to interpret its dynamics. In contrast to the steady-state case, now the rate at which excitons arrive in the emitting QD, as well as the spin-flip probability $p_0$, become important. We determine the former from biexciton-exciton intensity ratio, using the aforementioned exciton dynamics rate equation model. The latter enters model calculation as a free parameter. The calculation result, presented in Fig. 3c, describes well the experimental values, assuming $p_0$~=~$0.1$. 
It is also important to note that in our case, the photocreated excitons are well defined and no free photocreated carriers are expected to contribute to the spin relaxation. We excite resonantly excitons in the neighboring QD with an excitation power sufficiently low to make any nonresonant processes negligible. Therefore, in contrast to the ref \cite{besombesPRB08}, we do not consider the spin exchange with individual carriers surrounding the QD.

To conclude, we demonstrated that the information can be written in the spin state of a single Mn ion in a QD in an orientation process, exploiting exciton spin transfer from a neighbor non-magnetic QD. The orientation time varied between 20 ns and 100 ns over the used range of excitation power. We determined a storage time of information in the Mn spin. It is enhanced by application of a moderate static magnetic field and reaches hundreds of microseconds in the dark.  

\begin{acknowledgments}
This work was partially supported by the Polish Ministry of
Science and Higher Education as research grants in years 2006-2011,
and by the 6th Research Framework
Programme of EU (contract MTKD-CT-2005-029671).
\end{acknowledgments}

\end{document}